\documentclass[preprint,showkeys,secnumarabic,amsfonts,showpacs,amsmath,amssymb]{revtex4}
\usepackage[dvips]{color}
\usepackage{epsfig}
\usepackage{amsmath}
\usepackage{graphicx}
\def\Box{\hbox{$\rlap{$\sqcup$}\sqcap$}}

\begin{document}

\title{Cosmological observations in non-local $F(R)$ cosmology }

\author{H. Farajollahi}  \author{F. Tayebi}  \author{F. Milani}  \author{M. Enayati}
\affiliation{Department of Physics,
University of Guilan, Rasht, Iran}

\begin{abstract}
\noindent \hspace{0.35cm} In this article in a generalization of our previous work, \cite{far-mil}, we investigate the dynamics of the non-local $F(R)$ gravity
 after casting it into local form. The non-singular bouncing behavior and quintom model of dark energy are achieved without involving negative kinetic energy fields.
 Two cosmological tests are performed to constrain the model parameters. In case of phantom crossing the distance modulus predicted by the model best-fits the observational data. In comparison with the CPL parametrization for drift velocity, the model in some redshift intervals is in good agreement with the data.
\end{abstract}

\maketitle

\section{Introduction}\label{s:intro}

The cosmological observations have provided increasing evidence that our universe is undergoing a late-time
 acceleration expansion \cite{Tzi07}. To explain this observation, a variety of theories are put forward to introduce a
 new fluid called dark energy, which possesses a negative enough pressure \cite{Tot97}. According to the observational
 data from the Type Ia Supernovae \cite{Rie04} and WMAP satellite \cite{Jas05}, our universe is spatially flat and consist approximately of \%25
  dark matter and \%75 dark energy. From the simplest models for dark energy as cosmological constant to the recent proposed models such as quintessence \cite{Ben03}, phantom \cite{Cal02}, or
combination of these two in a unified model named quintom \cite{Sad08} all try to explain the accelerating expansion universe.

An alternative approach to explain dark energy is the modification of general relativity (GR) \cite{Def20}-\cite{Car05}. Here, as a generalization of the work in \cite{far-mil}, we consider a non-local modified gravity, where in addition an arbitrary function of Ricci scalar, $F(R)$, is added in the action, with the aim to explain the dark energy responsible for current universe acceleration. For cosmic acceleration, the equation of state $(EOS)$ parameter is less that $-1/3$, while for quintessence, Chaplygin gas \cite{Kam01} and holographic dark energy
models \cite{Li04}, $\omega$ always stays bigger than $-1$ and for the phantom  models is smaller than $-1$  \cite{Cai07}. For a dynamical dark energy, the EoS parameter evolves from $\omega > -1$ to $\omega< -1$ (or vice versa) and is called "the phantom divide crossing".

In particular, the quintom model of \cite{Fen05} were initially proposed to obtain a model of dark energy with the $\omega>-1$ in the past and
$\omega <-1$ at present. The model can be viewed as dynamical model for dark energy with
the feature that its EoS parameter can smoothly cross over the cosmological constant barrier $\omega =-1$. To construct a quintum model
it is necessary to add extra degrees of freedom with un-conventional features to the conventional single field
theory if we expect to realize a viable model in the framework of gravity theory \cite{Zha07}. Our model involves two scalars with one being Quintessence-like and another Phantom-like. The model provides a bouncing cosmology which allows us to avoid the problem of the initial singularity.

The paper is arranged as follows: in the next section we establish our model and give the exact solution of the scalar-tensor field equations. We also analytically investigate the conditions for the EoS parameter in the model to across $-1$  In section three, by using the field equations from the previous section we obtain numerical solutions for the cosmological scale factor, the EoS, Hubble and deceleration parameters. In Section four, we perform two cosmological tests to validate our model with the observational data. A summary is presented in section five.

\section{The model}

In this section we consider local formulation of
non-local $ F(R)$ gravity with the action given
by {\cite{Noj08}},
\begin{eqnarray}\nonumber
S=\int{d^{4}x\sqrt{-g}\left[\frac{1}{2}\left\{R(1+f(\phi)-\psi)-\partial_
{\mu}\psi\partial^{\mu}\phi\right\}+F(R)\right]},
\end{eqnarray}
\begin{eqnarray}
\label{ac}
\end{eqnarray}
where $f$ is a function of scalar field
 $\phi$, $F$ is a function of scalar curvature $R$ and $\psi$ is a second scalar field in the model. Variation of Eq. (\ref{ac}) with respect to the metric tensor
$g_{\mu\nu}$ gives,
\begin{eqnarray}\label{T}
0&=&\frac{1}{2}g_{\mu\nu}\left\{R(1+f-\psi)-\partial_{\rho}\psi\partial^{\rho}\phi\right\}-R_{\mu\nu}(1+f-\psi)\nonumber\\
&+&\frac{1}{2}(\partial_{\mu}\psi\partial_{\nu}\phi+\partial_{\mu}\phi\partial_{\nu}\psi)
-(g_{\mu\nu}\Box-\nabla_{\mu}\nabla_{\nu})(f-\psi)\nonumber\\
&-&2(g_{\mu\nu}\Box-\nabla_{\mu}\nabla_{\nu})F'-2R_{\mu\nu}F'+g_{\mu\nu}F,
\end{eqnarray}
where $F'=\frac{d F(R)}{d R}$. In a FRW cosmological model, for only time dependent $\phi$ and $\psi$, by invariance of the action under changing fields and
vanishing variations at the boundary, the equations of motion for
scalar fields $\phi$ and $\psi$ become,
\begin{eqnarray}
0&=&\ddot{\phi}+3H\dot{\phi}+R,\label{EoS1}\\
0&=&\ddot{\psi}+3H\dot{\psi}-R f^{(1)},\label{EoS2}
\end{eqnarray}
where $f^{(n)}=\frac{d^{n}f(\phi)}{d\phi^{n}}$, $R=12H^2+6\dot{H}$ and $H$ is Hubble parameter. Also, the $00$ and $ii$ components of equation (\ref{T}) have the following form
\begin{eqnarray}
0&=&-3H^2(1+f-\psi)+\frac{1}{2}\dot{\psi}\dot{\phi}-3H(f^{(1)}\dot{\phi}-\dot{\psi})-F\nonumber\\
&+&6(H^2+\dot{H})F'
-36(4H^2\dot{H}+H\ddot{H})F'',\label{f1}\\
0&=&(2\dot{H}+3H^2)(1+f-\psi)-2(\dot{H}+3H^2)F'+\frac{1}{2}\dot{\psi}\dot{\phi}\nonumber\\
&+&(\frac{d^2}{dt^2}+2H\frac{d}{dt})(f-\psi+2F')+F
\cdot\label{f2}
\end{eqnarray}\label{f3}
Equations (\ref{f1}) and (\ref{f2}) can be rewritten as
\begin{eqnarray}
3H^2&=&\{\frac{1}{2}\dot{\psi}\dot{\phi}-3H(f^{(1)}\dot{\phi}-\dot{\psi})-F
+6(H^2+\dot{H})F'\nonumber\\
&-&36(4H^2\dot{H}+H\ddot{H})F''\}/(1+f-\psi),\label{f3}\\
-2\dot{H}&-&3H^2=\{(\frac{d^2}{dt^2}+2H\frac{d}{dt})(f-\psi+2F')
+\frac{1}{2}\dot{\psi}\dot{\phi}\nonumber\\
&-&2(\dot{H}+3H^2)F'+F\}/(1+f-\psi)\cdot\label{f4}
\end{eqnarray}
In comparison with the standard Friedman equations, one can treat the model as standard model with the effect of the gravity modification gives rise to a new energy density and pressure for the Friedman equations. So the right hand side of the equations (\ref{f3}) and (\ref{f4}) are
\begin{eqnarray}
\rho_{eff}&=&\{\frac{1}{2}\dot{\psi}\dot{\phi}-3H(f^{(1)}\dot{\phi}-\dot{\psi})-F
+6(H^2+\dot{H})F'\nonumber\\
&-&36(4H^2\dot{H}+H\ddot{H})F''\}/(1+f-\psi),\label{f5}\\
p_{eff}&=&\{\frac{1}{2}\dot{\psi}\dot{\phi}+(\frac{d^2}{dt^2}+2H\frac{d}{dt})(f-\psi+2F')+F\nonumber\\
 &-&2(\dot{H}+3H^2)F'\}/(1+f-\psi)\cdot\label{f6}
\end{eqnarray}
After some algebraic calculation, we can read the effective energy density and pressure from the above equations as,
\begin{eqnarray}\label{rho}
\rho_{eff}&=&\frac{1}{\mathcal{A}^2-4F'^2}\{\frac{1}{2}\dot\phi\dot\psi(\mathcal{A}-4F')-3H\dot{\mathcal{B}}(\mathcal{A}+F')\nonumber\\
&-&F(\mathcal{A}+2F')-3\ddot{\mathcal{B}}F'\},
\end{eqnarray}
\begin{eqnarray}\label{p}
p_{eff}&=&\frac{1}{\mathcal{A}^2-4F'^2}\{\frac{1}{2}
\dot{\psi}\dot{\phi}\mathcal{A}+F\mathcal{B}+2(\mathcal{A}+\frac{5}{2}F')H\dot{\mathcal{B}}\nonumber\\
&+&(\mathcal{A}+F')\ddot{\mathcal{B}}\}.
\end{eqnarray}
where $\mathcal{A}\doteq 1+f-\psi$ and $\mathcal{B}\doteq \mathcal{A}+2F'$ and $p_{eff}=\omega_{eff}\rho_{eff}$. From Eqs. (\ref{f1}) and (\ref{rho}) we obtain,
\begin{eqnarray}\label{h}
H=\frac{1}{2(\mathcal{A}^2-4F'^2)}\left(-\mathcal{C}\pm\sqrt{\mathcal{C}^2-\frac{4}{3}\left(\mathcal{A}^2-4F'^2\right)^2
(\mathcal{D}+3\mathcal{E}F')}\right)
\end{eqnarray}
where $\mathcal{C}\doteq(f^{(1)}\dot{\phi}-\dot{\psi}+2\frac{dF'}{dt})(\mathcal{A}+F')$, $\mathcal{D}\doteq-\frac{1}{2}\dot{\psi}\dot{\phi}(\mathcal{A}-4F')+F\mathcal{B}$ and $\mathcal{E}\doteq f^{(2)}\dot{\phi}^2+f^{(1)}\ddot{\phi}-\ddot{\psi}+2\frac{d^2F'}{dt^2}$. Using Eqs. (\ref{f4}), (\ref{rho}), (\ref{p}) and (\ref{h}), one can find,
\begin{eqnarray}\label{Hdot}
\dot{H}=\frac{-1}{2\mathcal{B}}\{\dot{\psi}\dot{\phi}-H\dot{\mathcal{B}}+\ddot{\mathcal{B}}\}\cdot
\end{eqnarray}
Also from Eq. (\ref{Hdot}), we have,
\begin{eqnarray}\label{hddot}
\ddot{H}&=&\frac{-1}{2\left(\mathcal{B}-6f^{(1)}\right)}\big\{2\left(\frac{d^3}{dt^3}-H\frac{d^2}{dt^2}+\dot{H}\frac{d}{dt}\right)F'\nonumber\\
&+&\ddot{\phi}\left(\dot{\psi}-4Hf^{(1)}+3f^{(2)}\dot{\phi}\right)
+\ddot{\psi}(\dot{\phi}+4H)\nonumber\\
&+&\dot{\phi}\left[f^{(3)}\dot{\phi}^2-f^{(2)}
\left(H\dot{\phi}+6\dot{H}+12H^2\right)\right]\nonumber\\
&+&2\dot{H}\left(\dot{\psi}-f^{(1)}(24H+1)\right)\big\},\nonumber
\end{eqnarray}
where $\mathcal{B} \neq 6f'$ is a constraint on $\ddot{H}$. To explore the possibility of the effective EoS parameter, $\omega_{eff}$, crosses
$-1$, the condition, $\ddot{H}=-\dot{\omega}_{eff}\rho)|_{\omega_{eff}\rightarrow -1}\neq 0$ has to be satisfied.

\section{Numerical Results}

A numerical solution for the corresponding equations for EoS parameter, $\omega_{eff}$, the Hubble parameter, $H$, and the scale factor $a$ in our model is presented in this section. In here, we choose $t=0$ to be the bouncing point. Our solution for $H(t)$, Eq. (\ref{h}), and consequently $a(t)$ provides a dynamical universe with contraction for $t<0$, bouncing at $t=0$ and then expansion for $t>0$. The above analysis clearly can be seen in the numerical calculation given in Fig. 1. It also shows that at the bouncing point $t=0$, the scale factor does not vanish which means the universe avoid singularity faced in the usual Einstein cosmology.\\

\begin{figure}
\includegraphics[scale=.3]{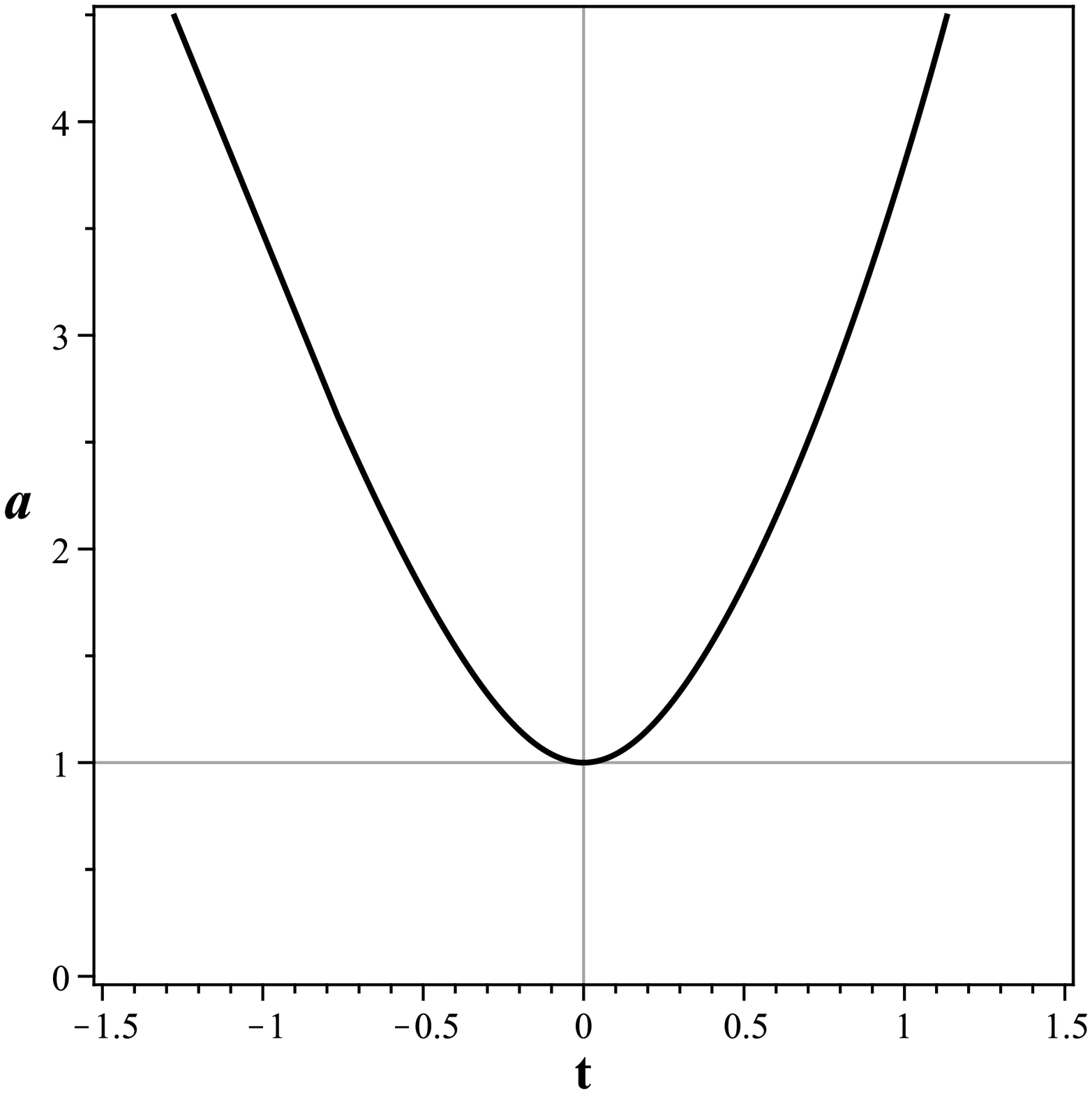}\hspace{.5 cm}\includegraphics[scale=.3]{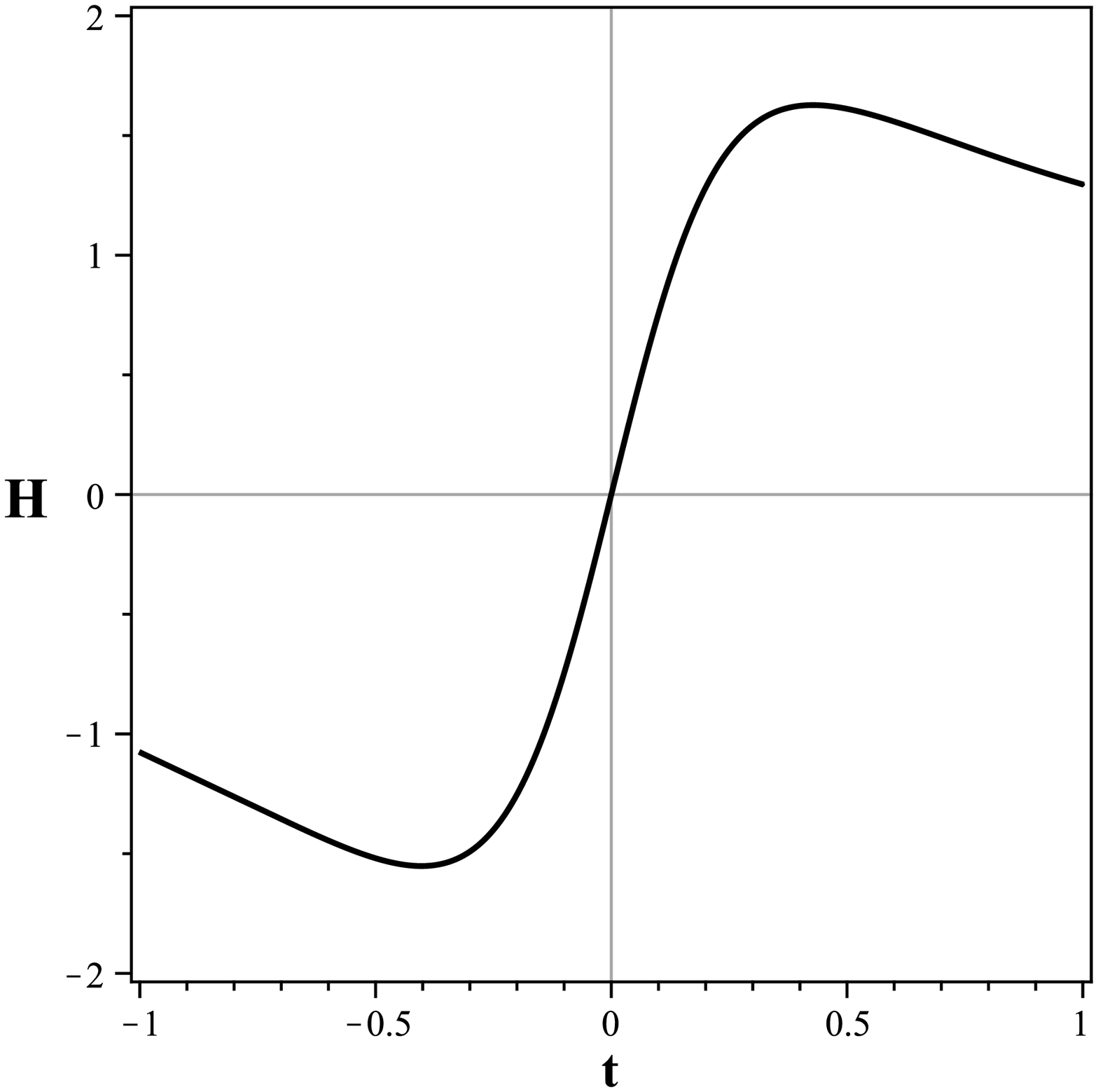}\\
Fig. 1: Plots of the evolution of scale factor $a(t)$ and Hubble parameter $H(t)$\\
 at the bounce for $F(R)=R^2$, ${\it f=f_{0}}\,{{\rm e}^{b\phi}}$, $f_{0} = -0.5$ and $b = 5$. Initial values are\\
 $\phi(0)=-5$, $\dot{\phi}(0)=15$, $\psi(0)=15$ , $\dot{\psi}(0)=10$, $\dot{H}(0)=8$ and $\ddot{H}(0)=1$.
\end{figure}

Our solutions for $\rho_{eff}$ and $p_{eff}$, Eqs. (\ref{rho}), (\ref{p}) shows that the effective EoS parameter $\omega_{eff}$ approaches negative values less than $-1$ at about $t=0$ and crosses $-1$ around this point as shown in Fig. 2. It also shows that in this model the EoS parameter crosses $-1$ line from
 $\omega<-1$ to $\omega>-1$ which is supported by observations  \cite{Guo05}.

In some sense this model is similar to quintom dark energy models that consist two quintessence and phantom fields \cite{Noz09}. A detailed examination on the necessary conditions requires
for a successful bounce shows that during the contracting phase, the scale
factor $a(t)$ is decreasing, i.e., $\dot{a} < 0$, and in the
expanding phase we have $\dot{a} > 0$. At the bouncing point,
$\dot{a} = 0$, and so around this point $\ddot{a} > 0 $ for a
period of time. Equivalently in the bouncing cosmology the Hubble
parameter $H$ runs across zero from $H < 0$ to $H > 0$ and $H = 0$
at the bouncing point. A successful bounce requires that the following condition should be satisfied around bouncing
point,
\begin{eqnarray}\label{hdot1}
\dot{H}=-\frac{1}{2M^{2}_{p}}(1+\omega)\rho>0.
\end{eqnarray}
From Fig. 1 and 2, we see that at $t\rightarrow 0$, $\omega_{eff}<-1$ and $\dot{H}$ is positive which satisfies the above condition. Also from Fig. 2, around $t=0$, for deceleration parameter, $q(t)$, we have a large acceleration which followed by a period of deceleration and a second acceleration. It should be noted that our model violate the strong energy condition and
the null energy condition (NEC), $(\rho_{eff}+p_{eff})<0$ for a short period around the bounce point and maybe physically
unstable. Moreover, after the
bounce the equation-of-state (EoS) parameter is able to transit from $\omega_{eff} < -1$
to $\omega_{eff} >-1$ and then connect with normal expanding history.\\

\begin{figure}
\includegraphics[scale=.3]{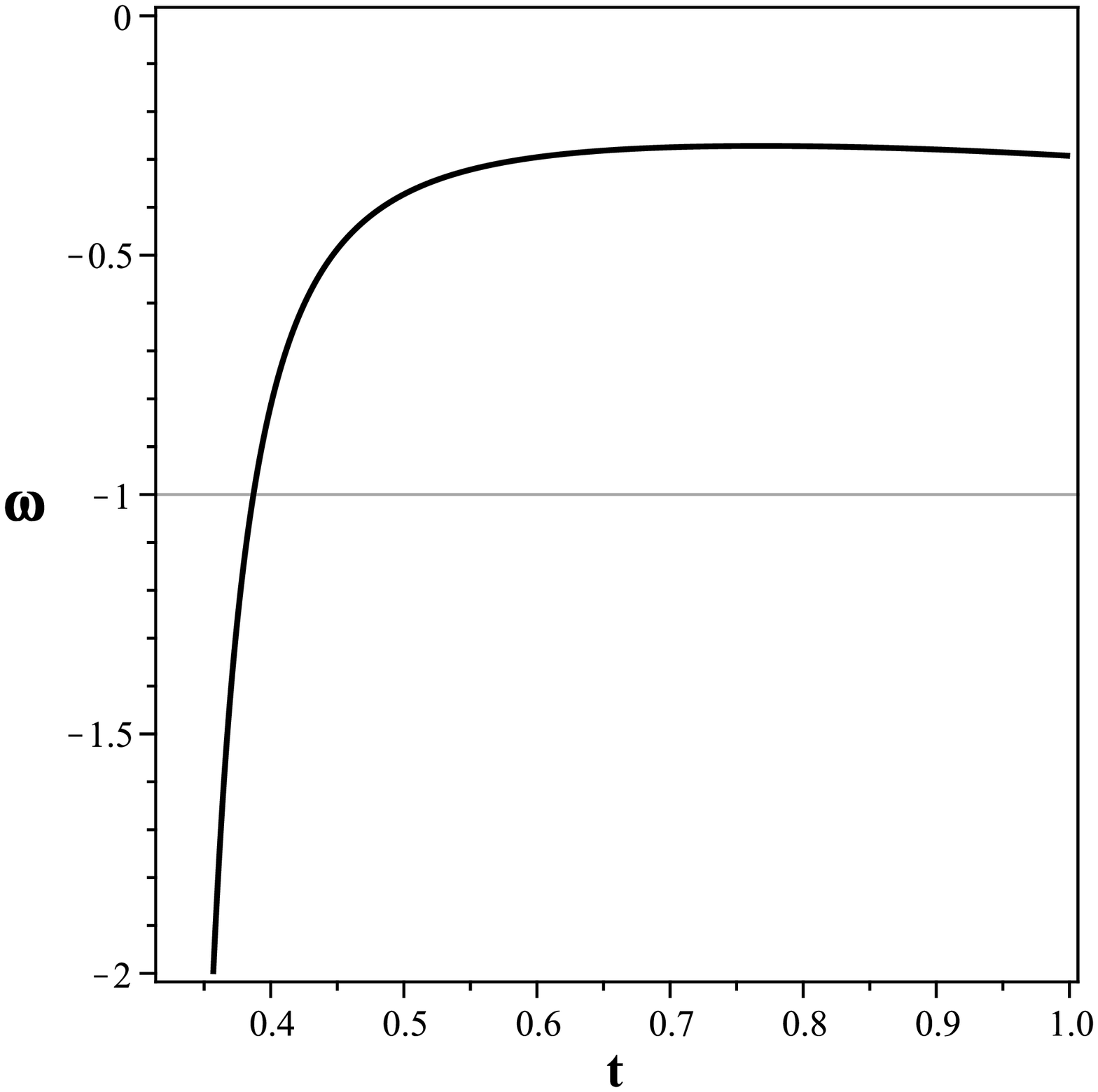}\hspace{.5 cm}\includegraphics[scale=.3]{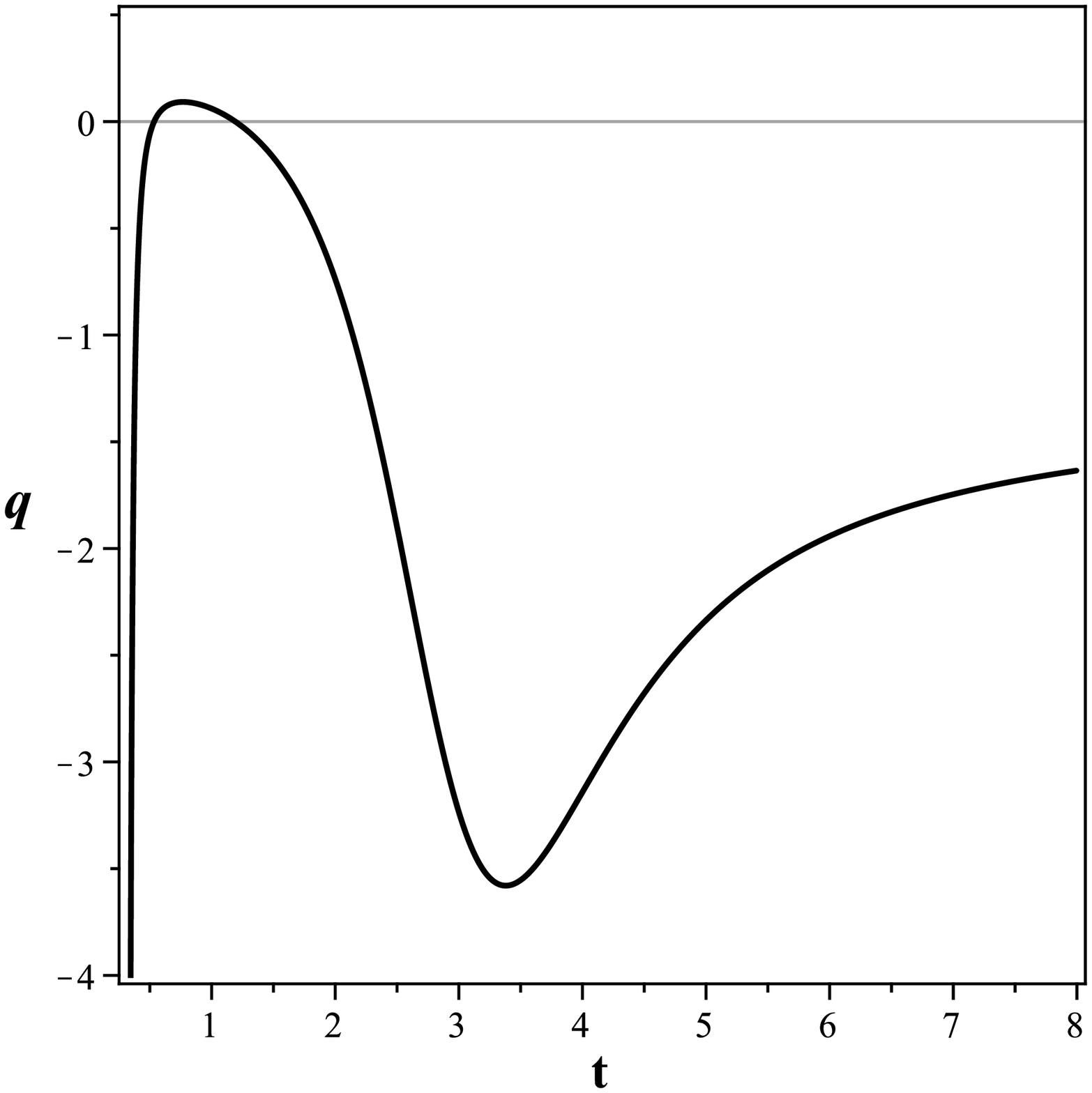}\\
Fig. 2: plots of EoS parameter $\omega(t)$ and deceleration parameter, $q(t)$, for $F(R)=R^2$, \\${\it f=f_{0}}\,{{\rm e}^{b\phi\left( t \right) }}$,
$f_{0} = -0.5$ and $b = 5$. Initial values are $\phi(0)=-5$, $\dot{\phi}(0)=15$, $\psi(0)=15$,\\
$\dot{\psi}(0)=10$, $\dot{H}(0)=8$ and $\ddot{H}(0)=1$.
\end{figure}

From Fig. 3, it is shown that the numerical solutions of the proposed model can reproduce the expected red-shift behaviors of the EoS parameter $\omega_{eff}(z)=-1$ at $z = 0.29$ \cite{Zha07a}.\\

\begin{figure}
\includegraphics[scale=.3]{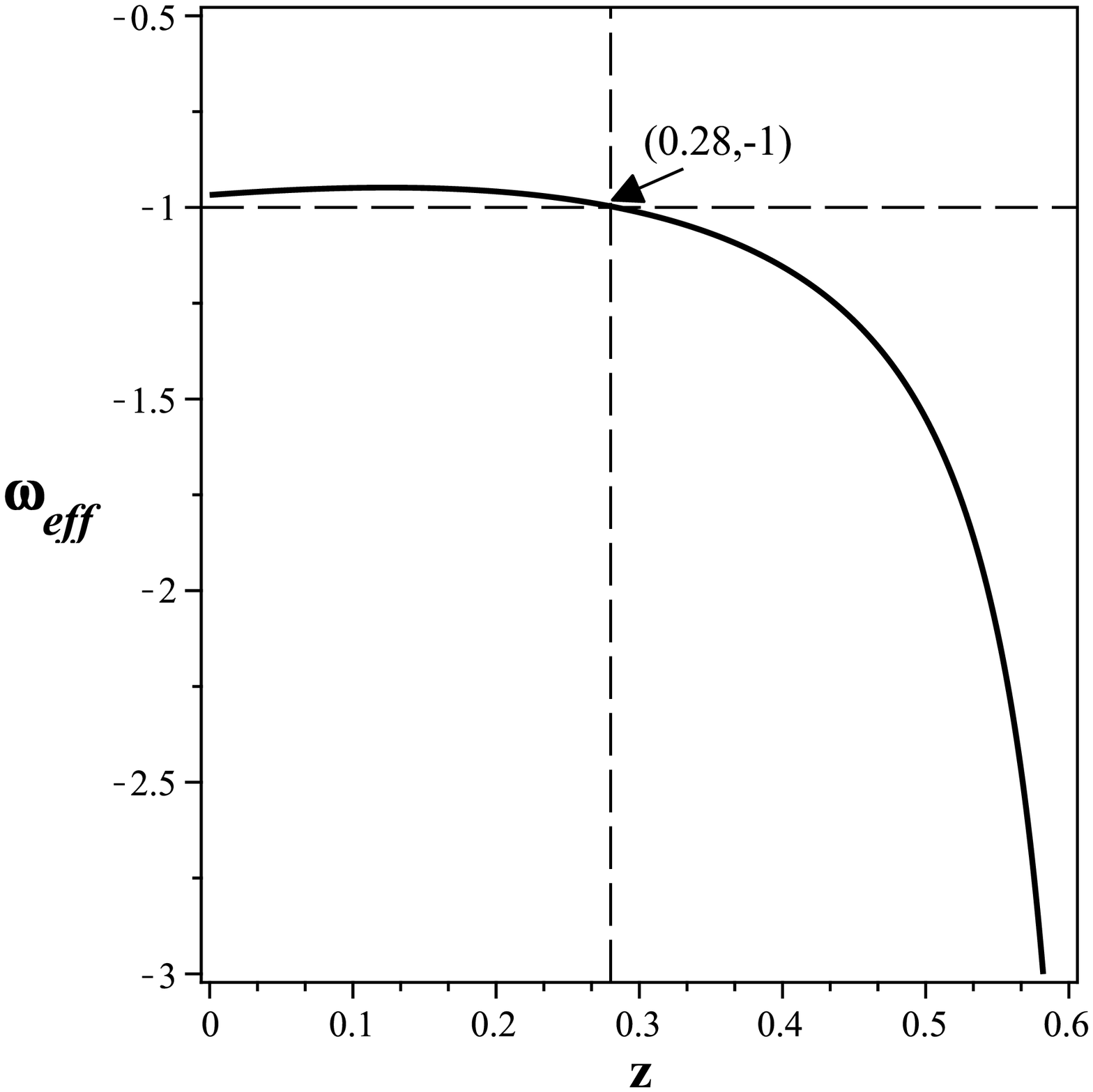}\hspace{.5 cm}\\
Fig. 3: plot of EoS parameter $\omega_{eff}(z)$ for $F(R)=R^2$, ${\it f=f_{0}}\,{{\rm e}^{b\phi\left( t \right) }}$,
$f_{0} = -0.5$ and $b = 5$.\\ Initial values are $\phi(0)=-5$, $\dot{\phi}(0)=15$, $\psi(0)=15$,
$\dot{\psi}(0)=10$, $\dot{H}(0)=8$ and $\ddot{H}(0)=1$.
\end{figure}

\section{Cosmological Tests}
We now examine our model with the observational data using the following cosmological tests \cite{Lis08}--\cite{Che01}.

\subsection{The difference in the distance modulus, $\mu(z)$}

The difference between the absolute and
apparent luminosity of a distance object is given by, $\mu(z) = 25 + 5\log_{10}d_L(z)$ where the Luminosity distance quantity, $d_L(z)$ is given by
\begin{equation}\label{dl}
d_{L}(z)=(1+z)\int_0^z{\frac{dz'}{H(z')}}.
 \end{equation}
In our model, from numerical computation one obtains $H(z)$ which can be used to evaluate $\mu(z)$. In Fig. 4, we
compare $\mu(z)$ in our model with the observational data. As can be seen the model for the given conditions of dynamical variables is in good agreement with the observational data.\\

\begin{figure}
\includegraphics[scale=.3]{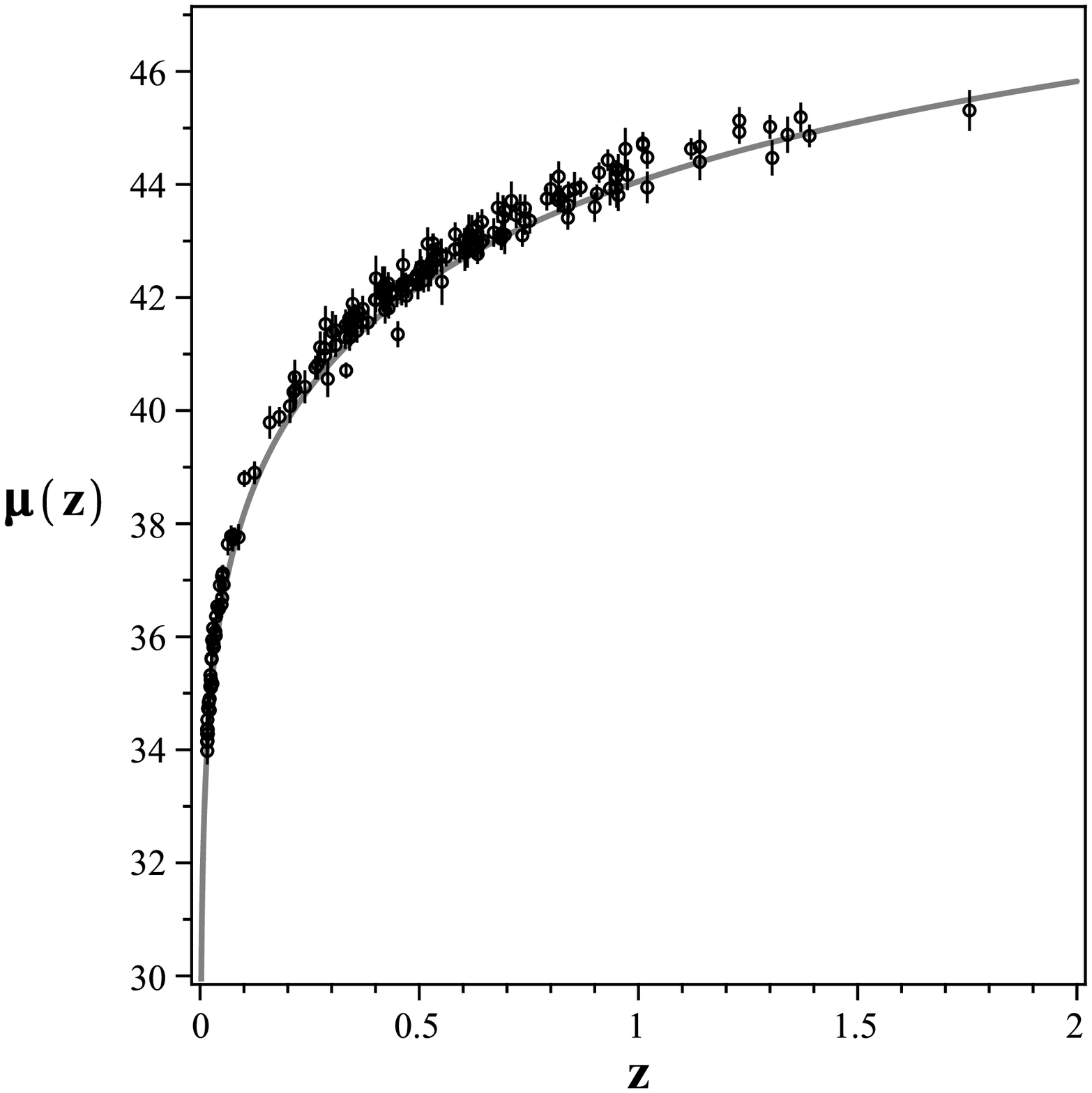}\hspace{0.1 cm}\\
Fig. 4: The graph of distance modulus $\mu(z)$ plotted as function of redshift, \\
in our model ( gray curve) in comparison with the observational data (the black data points).\\
 ICs. $\phi(0)=-0.5$, $\dot{\phi}(0)=0.13$, $\psi(0)=-0.6$,
$\dot{\psi}(0)=-0.2$, $\dot{H}(0)=0.1$ and $\ddot{H}(0)=1$.
\end{figure}

\subsection{CPL, CRD and our model}

Following~\cite{Che01}, in Chevallier-Polarski-Linder (CPL) parametrization model one can use linearly approximated EoS parameter,
\begin{eqnarray}\label{hdot1}
\omega_{cpl}\approx \omega_0-\frac{d\omega_{cpl}}{da}(a-1)=\omega_0+\omega_1\frac{z}{1+z},
\end{eqnarray}
where $\omega_0$ is current value of the EoS and $\omega_1=-\frac{d\omega_{cpl}}{da}$ is its running factor. Using the above equation we can find the following equation for Hubble parameter,
\begin{eqnarray}\label{Hr}
\frac{H(z)^{2}}{H^{2}_{0}} &=&\Omega_m(1+z)^3+(1-\Omega_m)(1+z)^{3(1+\omega_0+\omega_1)}\nonumber\\
    & \times&  \exp{\left[-3\omega_1(\frac{z}{1+z})\right]}.
 \end{eqnarray}
In CPL model the parametrization is fitted for values of $\omega_0$, $\omega_1$ and $\Omega_m$ corresponding to our model. On the other hand, the CRD can be extracted from
\begin{eqnarray}\label{dotz}
\dot{z}=(1+z)H_0-H(z),
\end{eqnarray}
which is known as Mc Vittie equation. This equation immediately leads to velocity drift,
\begin{eqnarray}\label{vdrift}
\dot{v}=cH_0-\frac{cH(z)}{1+z}.
\end{eqnarray}
By using equation (\ref{Hr}) in CPL model, the velocity drift with respect to the redshift can be obtained against observational data. Fig. 5, shows the velocity drift in terms of redshift against observational data for our model.\\

\begin{figure}
\includegraphics[scale=.3]{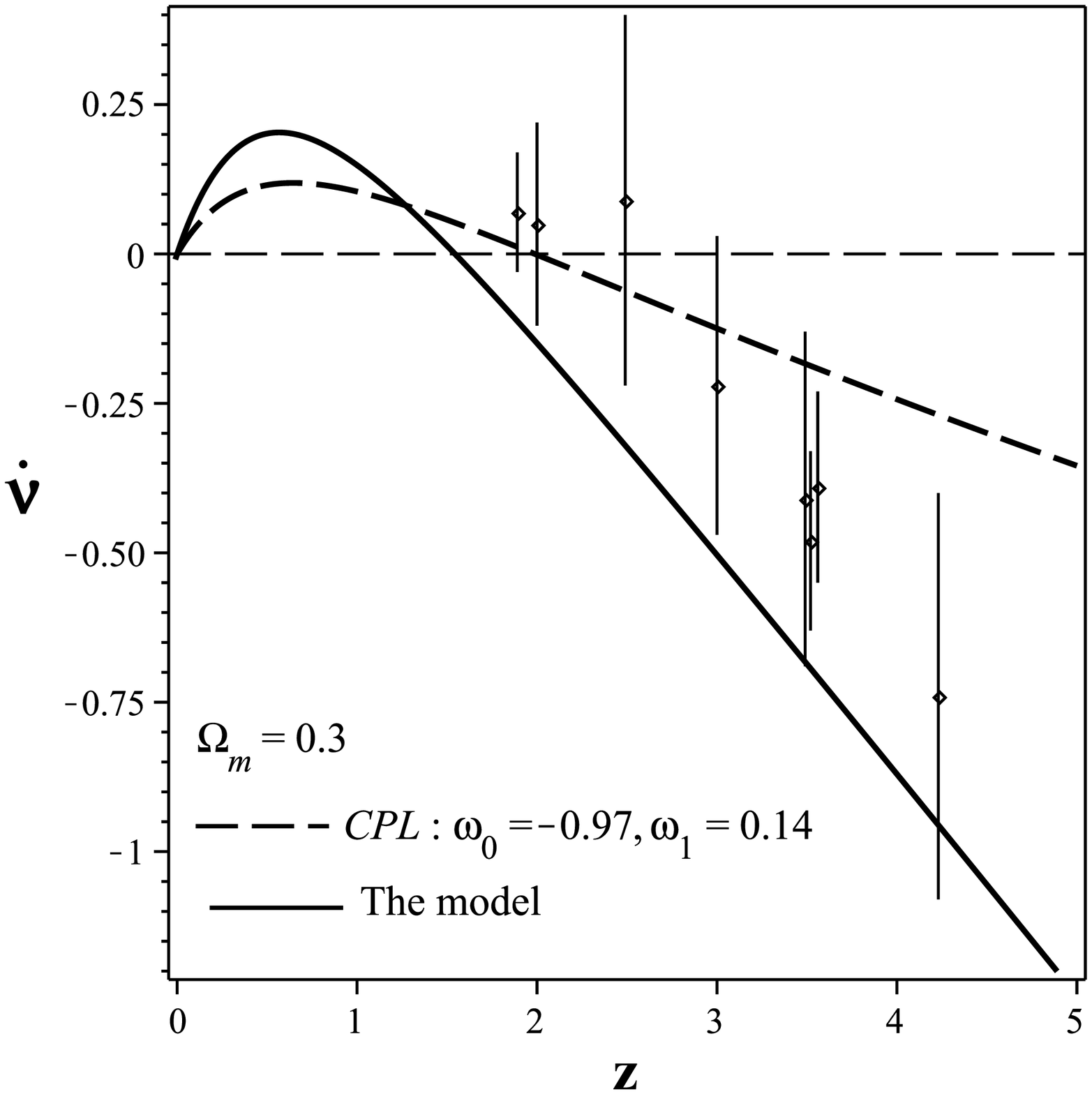}\\
Fig. 5: The graph of velocity drift plotted as function of redshift $z$ for both CPL \\parametrization and
our model.  ICs. $\phi(0)=-0.5$, $\dot{\phi}(0)=0.13$, $\psi(0)=-0.6$,\\
$\dot{\psi}(0)=-0.2$, $\dot{H}(0)=0.1$ and $\ddot{H}(0)=1$.
\end{figure}

From Fig. 5, a comparison between our model and CPL model with respect to the observational data is given. As can be seen that our model for the redshift $z$ greater than $3.5$ better fits the data than CPL parameter.

\section{Summary}

In this paper, we have studied the evolution of the gravitational fields both analytically and
numerically in the non-local $F(R)$ model in which the geometry partly modified by two scalar fields, one
of them a Quintessence-like field, the other a Phantom-like field.

We discussed a bouncing non-singular cosmology, with an initial contracting phase which lasts until to a non-vanishing
minimal radius is reached and then smoothly transits into an expanding phase which provides a possible solution
to the singularity problem of Standard Big Bang cosmology, a problem which is not cured by scalar-field driven
inflationary models. The evolution of EoS parameter, hubble parameter, scale factor, and deceleration parameter is numerically obtained. The violations of the null energy condition required to get a bounce are obtained for the model, which allows a transition of the EoS parameter through the cosmological
constant boundary. In comparison with our previous work in \cite{far-mil}, we find that in the analytical discussion of the phantom crossing behavior of the EoS parameter,
we need also constrain the scalar fields and their first and
second derivatives. Moreover, we have also additional constraints on hubble parameter and its first and second derivatives. In \cite{far-mil}, with reconstruction of function $f( \phi)$,
we found that its behavior is similar to exponential function. Here, we are motivated to assume an exponential behavior for $f( \phi)$. Beside, such exponential functions have been known lead to interesting
physics in accelerated expanding cosmological models \cite{far-salehi}.

In comparison to our previous work \cite{far-mil}, also two cosmological tests are performed to validate our model. The distance modulus obtained from numerical computation fitted with the observational data with a phantom crossing behavior for the universe is predicted at redshift $z=0.29$. With regards to the assumed functionality for $F(R)$ and $f(\phi)$, and the crossing behavior for the model, the second test is performed. From numerical calculation, the drift velocity is obtained for the model and is compared with the observational data and CPL parametrization. The result shows that our model may better fit the data in comparison with CPL model in higher redshifts .

\nocite{*}
\bibliographystyle{spr-mp-nameyear-cnd}
\bibliography{biblio-u1}

\end{document}